\documentclass{article}

\usepackage{graphicx}

\title{A phase-resolved spectroscopic study
of the X-ray binary system 4U0115+63
based on Beppo-Sax observations.  }

\author{A.N. Baushev\thanks{Istituto di Asrofisica Spaziale e Fisica cosmica,
C.N.R., sezione di Palermo,
 Via Ugo La Malfa 153, 90146 Palermo, Italy\endgraf
Bogoliubov Laboratory of Theoretical Physics, Joint Institute for Nuclear Research;
141980 Dubna, Moscow Region, Russia\endgraf Email: abaushev@mx.iki.rssi.ru}}

\date{}

\begin{document}
\setcounter{topnumber}{1}
\maketitle

\abstract{We present a phase resolved spectral analysis of the
X-ray binary 4U0115+63 based upon Beppo-Sax satellite
observations. Three strong absorption lines have been detected at
all phases. We interpreted them as a cyclotron resonant
scattering. Existence of the fourth cyclotron line has been
confirmed at some phases. The cyclotron feature is found to be
strongly dependent on the phase of the source, while the continuum
part of the spectrum depends on the phase relatively weakly. The
cyclotron lines turned out to be very nonequidistant. The second
line is almost always deeper then the first one. We discussed
physical conditions in the active regions of the source, under
which a cyclotron feature similar to the observed one can appear.}

\section{Introduction.}

\begin{figure*}
\resizebox{0.8\hsize}{!}{\includegraphics[angle=270]{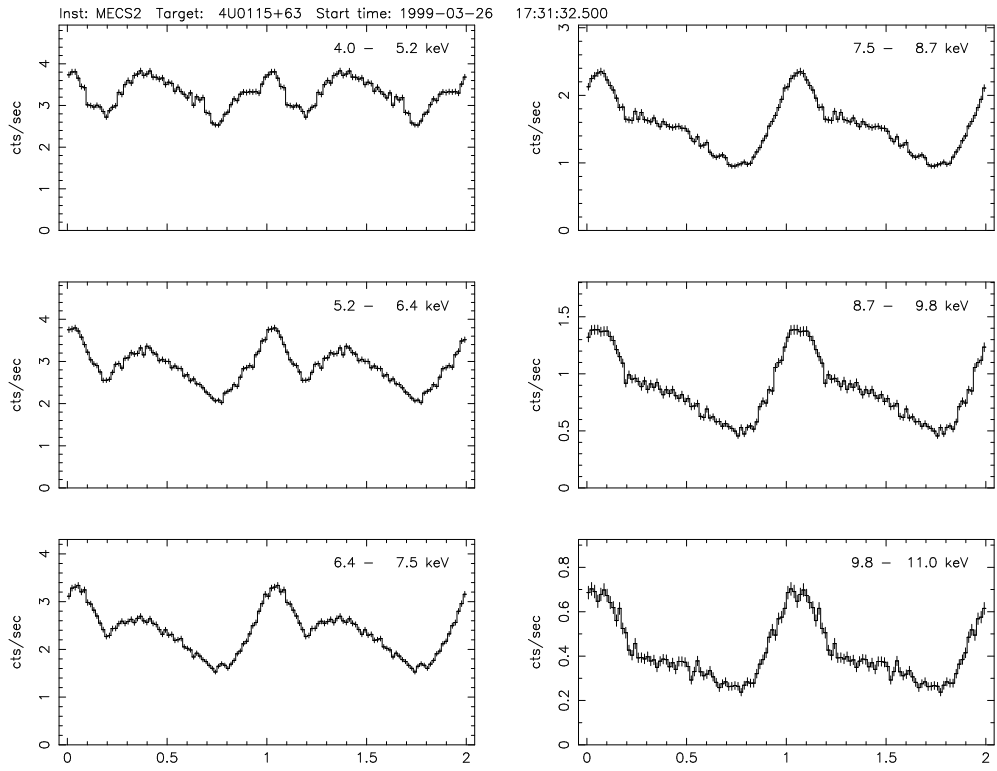}}
\caption{Phase profiles of the MECS2 data for the energy ranges
$4-11$~keV.} \label{fig1a}
\end{figure*}

\begin{figure*}
\resizebox{0.8\hsize}{!}{\includegraphics[angle=270]{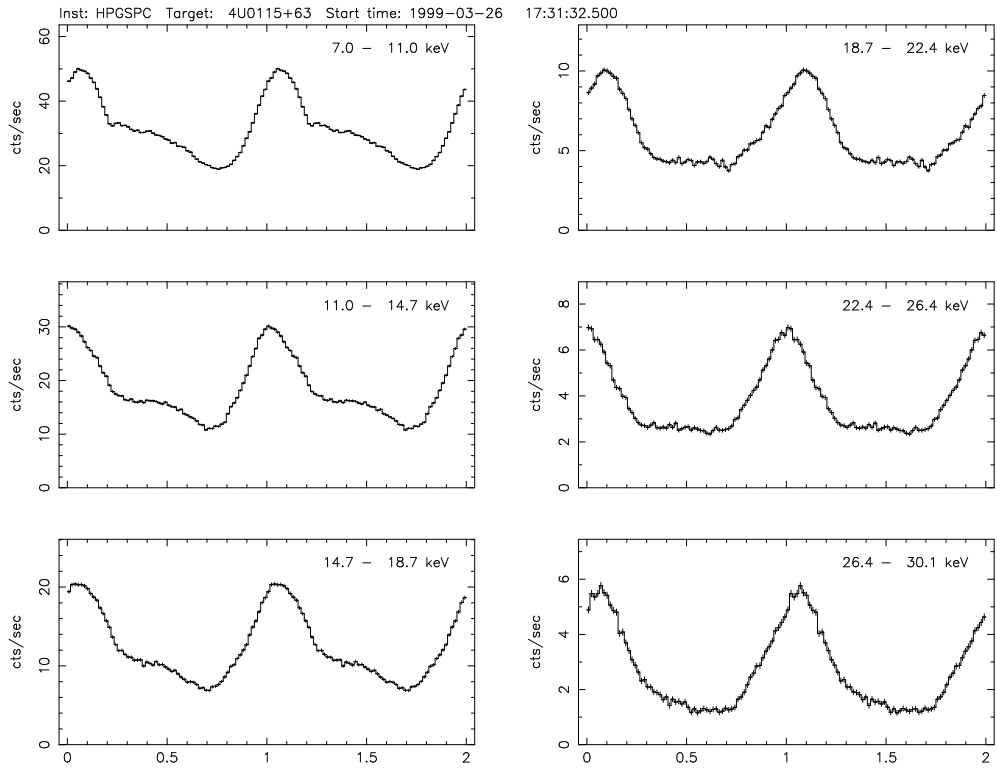}}
\caption{Phase profiles of the HPGSPC data for the energy ranges
$7-30$~keV.} \label{fig1b}
\end{figure*}

\begin{figure*}
\resizebox{0.8\hsize}{!}{\includegraphics[angle=270]{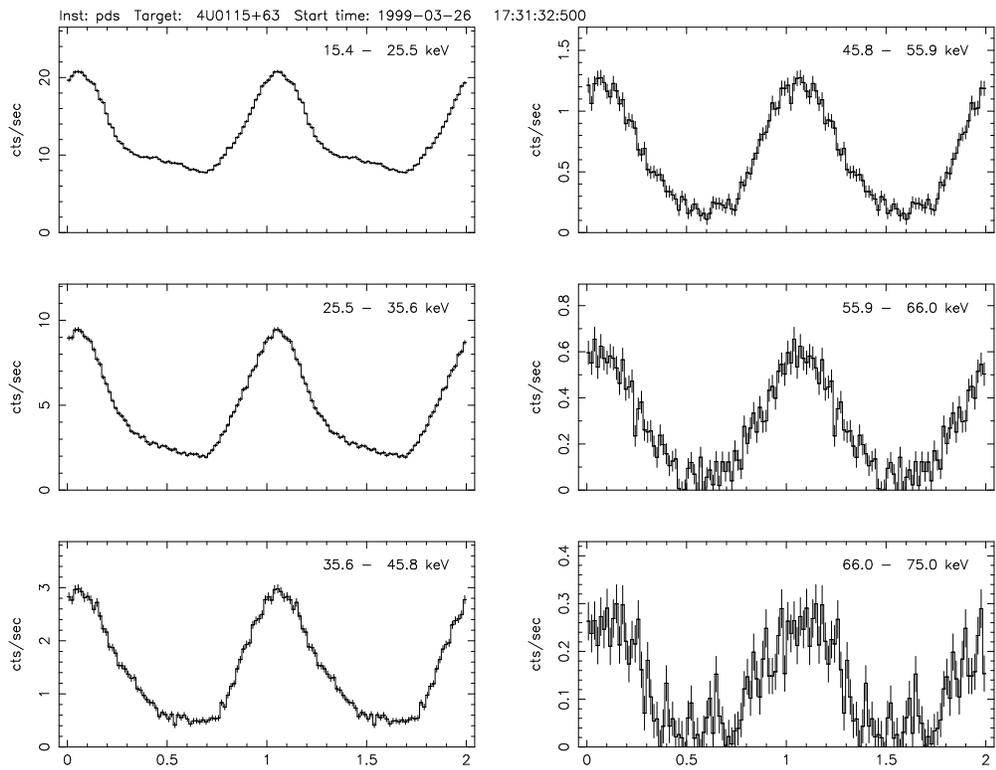}}
\caption{Phase profiles of the PDS data for the energy ranges
$15-75$~keV.} \label{fig1c}
\end{figure*}

\begin{figure*}
\resizebox{0.8\hsize}{!}{\includegraphics[angle=270]{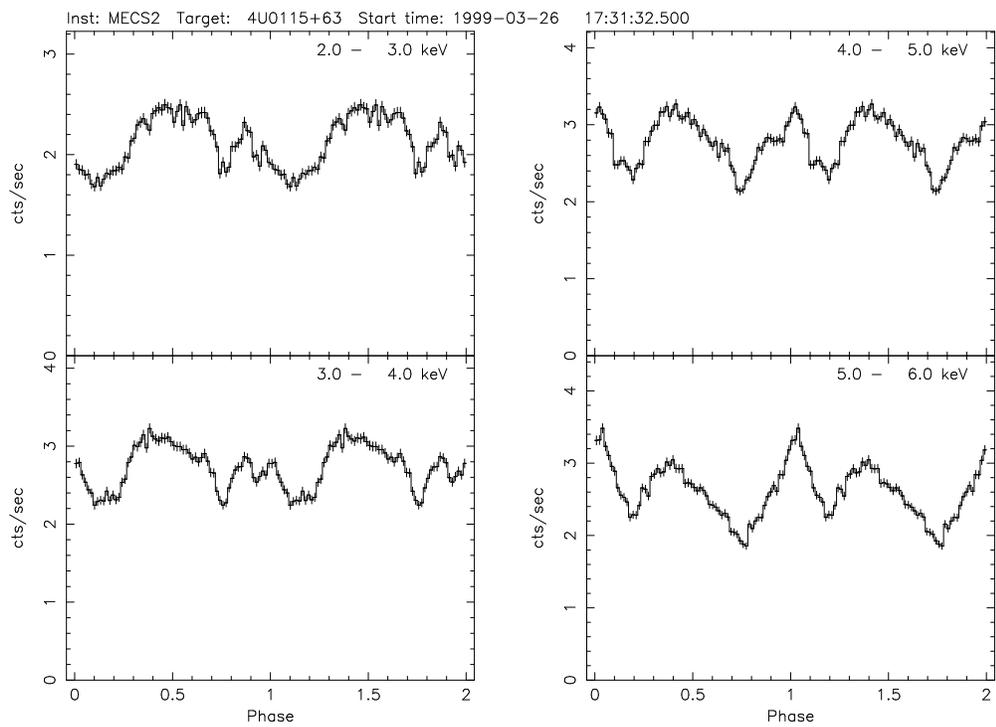}}
\caption{Phase profiles of the source 4U0115+63 for the energy
ranges $2-6$~keV.} \label{softdetail}
\end{figure*}

\begin{figure}[t]
\resizebox{0.8\hsize}{!}{\includegraphics[angle=270]{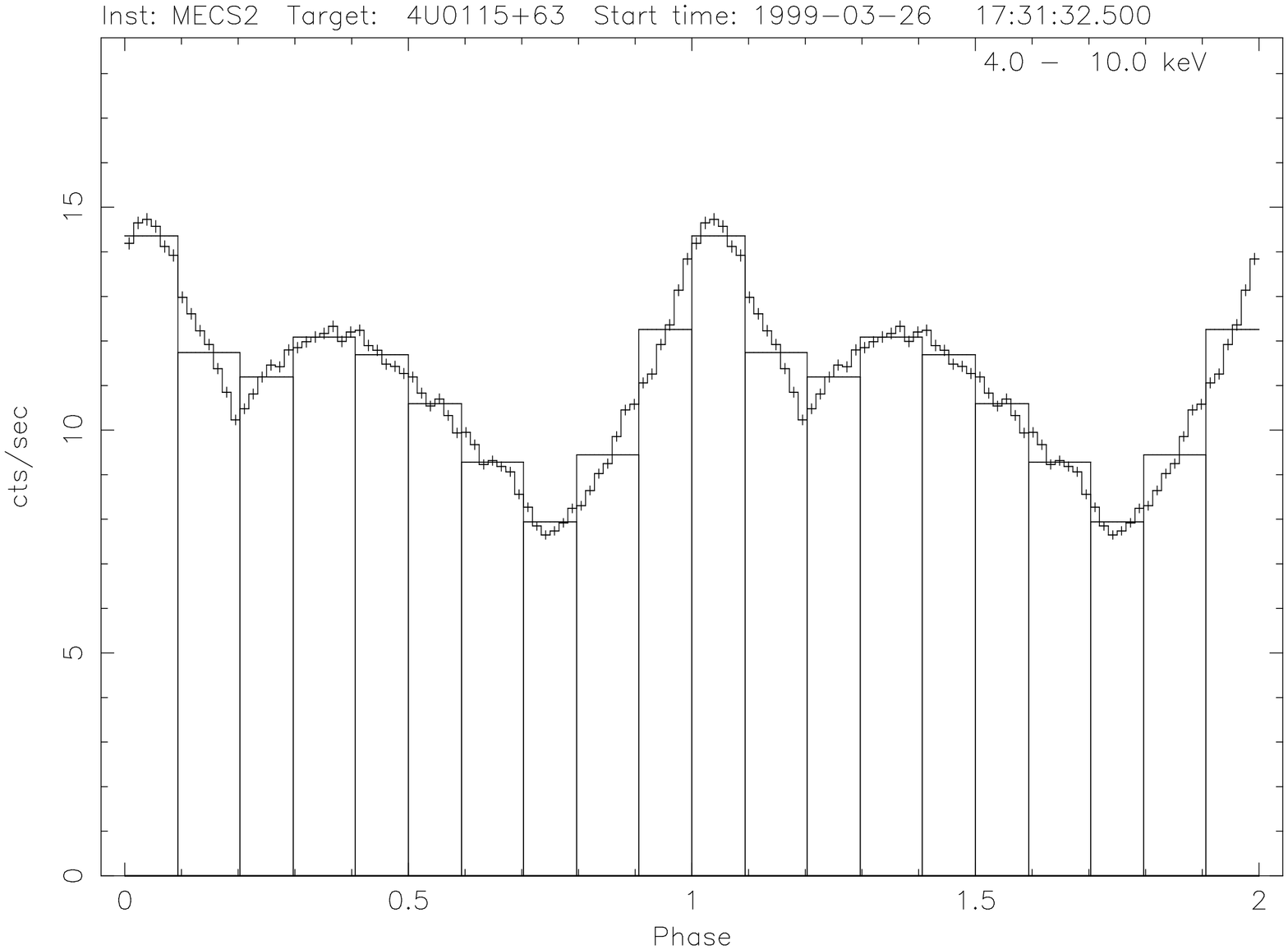}}
\caption{Pulse profile for the energy range $4-11$~keV. Phase
region selection is shown.} \label{fig09}
\end{figure}

4U0115+63 is a well-known transient massive X-ray binary source.
The orbital period of the binary system is $24.3$ days, the
eccentricity of the orbit is ${\it e}=0.34$ \cite{rappaport78}.
The optical companion of the source is $V635$ Cassiopeiae, an $O9$
star \cite{unger98}. X-ray outburst from 4U0115+63 has been
observed first with {\it UHURU} \cite{forman76}, and later with
{\it HEAO-1} \cite{rose79}, {\it CGRO/BATSE} \cite{bildsten97},
{\it RXTE} \cite{coburn02}, {\it Ginga} \cite{mihara04}, {\it
Beppo-SAX} \cite{santangelo99b}. The source shows pulsations with
the period $\sim$3.6 s \cite{rose79}.

A cyclotron resonant feature in the spectrum of 4U0115+63 has been
found out in 1979 \cite{wheaton79}. In 1983 the second cyclotron
harmonic was discovered \cite{white83}. Later the third and the
forth cyclotron absorption lines have been detected in the
spectrum of the source \cite{heindl99,santangelo99b}. Mihara et
al. fulfilled a phase-resolved spectroscopy of 4U0115+63 based
on {\it Ginga} observations \cite{mihara04}. They have found a
clear dependence of the parameters of the cyclotron feature on the
phase of the source and its luminosity.

By now 4U0115+63 is the only X-ray binary having four cyclotron
lines observed in its spectrum. This unique property  make the
source 4U0115+63 a touchstone for theories of the gyromagnetic
emission of X-ray pulsars. If only one weak line is visible in the
spectrum of a pulsar, usually it can be fitted well enough by a
very wide group of theoretical models. A big number of feature's
parameters in the case of 4U0115+63 could give as an opportunity
to choose from this abundance of models the appropriate one.

Here we report on the phase-resolved spectral analysis of
4U0115+63 based on Beppo-Sax satellite observations of the
source's outburst in 1999 March. We concentrated particulary on
the cyclotron feature and its evolution with the phase. Obtained
results allowed us to analyze the conditions in the active region
of the source and physical processes leading to the formation of
the cyclotron lines.

\section{Observations and spectral analysis.}
\subsection{Time analysis.}
This work is based upon the observations performed by Beppo-Sax
satellite on 1999 March, 26 (OP6714). Observation time was equal
$17^h 12.5^s$. Since the first cyclotron harmonic in the
spectrum of 4U0115+63 is near $\sim$12~{keV} \cite{white83,
santangelo99b} we used the data from the spectral interval from
2~{keV} to 75~{keV}. This energy range is covered by three
Beppo-Sax instruments: MECS2(2-11~{keV}), HPGSPC (7-30~{keV}),
PDS(15-75~{keV}).  More detail information about these instruments
can be found in \cite{boella97, manzo97, frontera97,
santangelo99b}.

\begin{figure}[t]
\resizebox{\hsize}{!}{\includegraphics[angle=270]{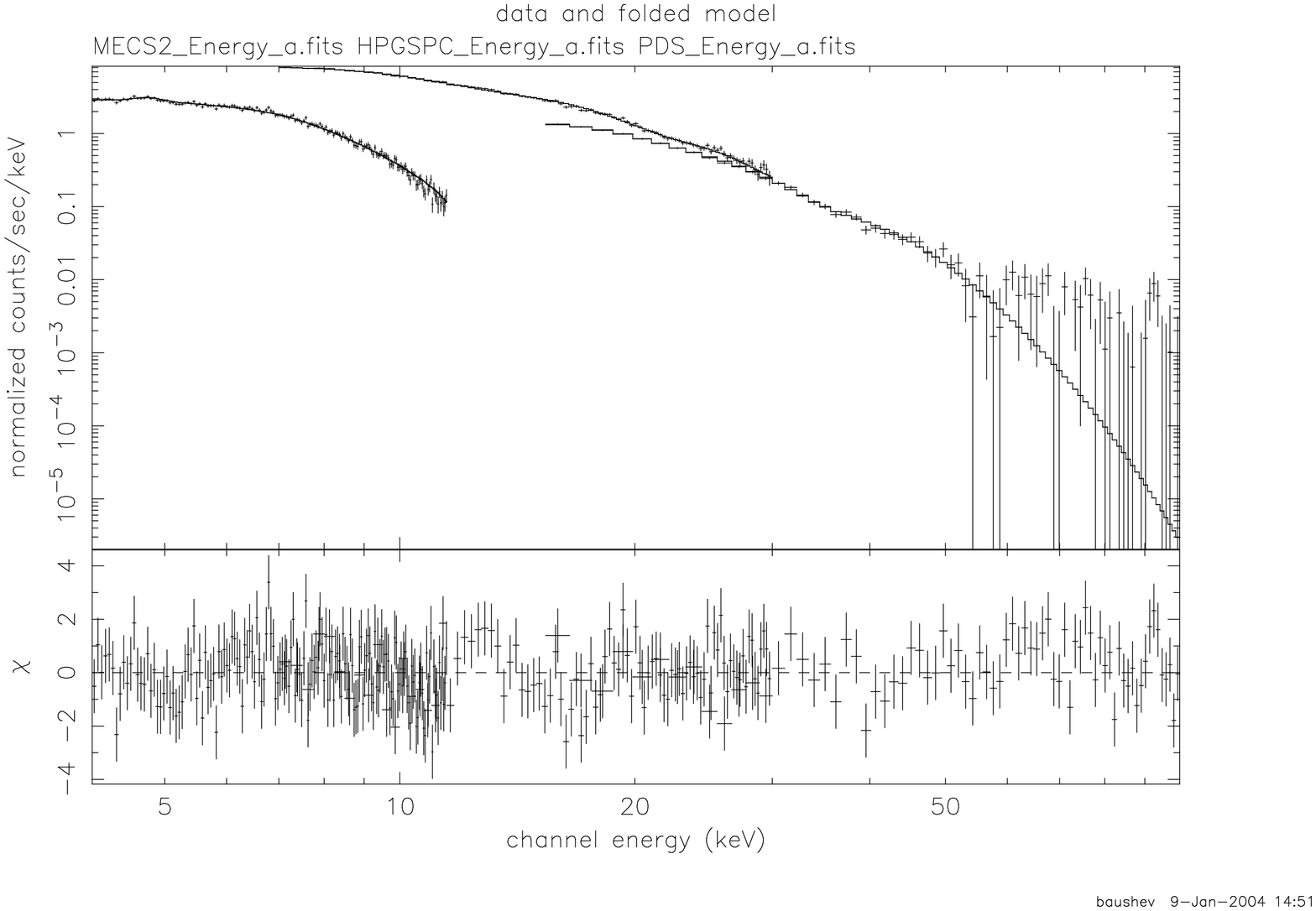}}
\caption{The spectrum of the pulsar at the phase {\em\bf 0,55}, in the soft peak.}
\label{fig2}
\end{figure}

\begin{figure}[t]
\resizebox{\hsize}{!}{\includegraphics[angle=270]{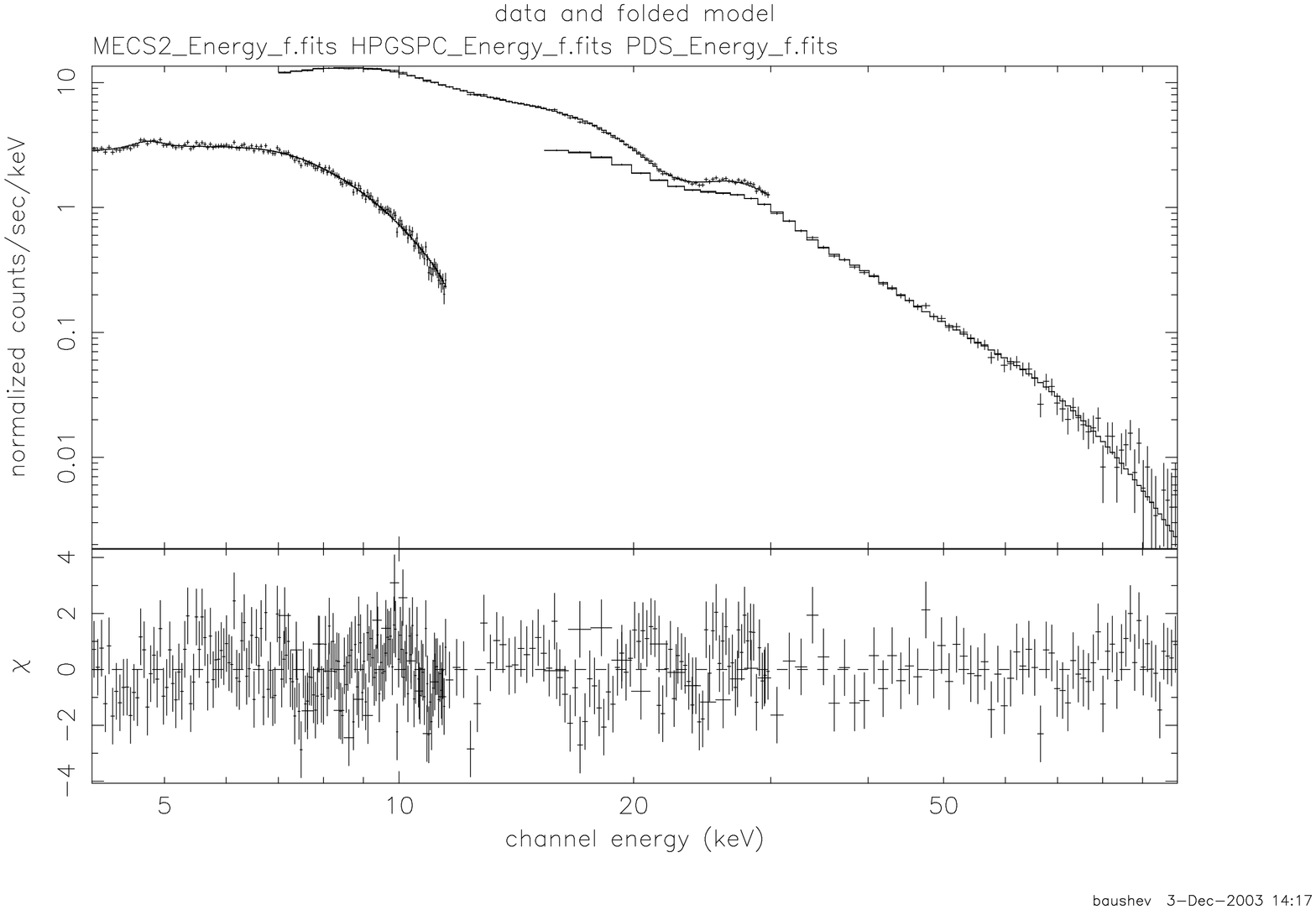}}
\caption{The spectrum of the pulsar at the phase {\em\bf 1,05}, in the center of the hard peak.}
\label{fig3}
\end{figure}

\begin{table}[b]
\begin{tabular}{cr@{$\pm$}lr@{$\pm$}l}
\hline
\hline
 Phase & \multicolumn{2}{c}{$a$}&\multicolumn{2}{c}{$T_1$}\\
\hline
0,45 & -0,673 & 0,074 & 5,48& 0,14\\
0,55 & -0,718 & 0,17 & 5,25 & 0,28\\
0,65 & -0,253 & 0,18 & 6,15 & 0,40\\
0,75 & -0,110 & 0,14 & 7,28 & 0,38\\
\hline
\end{tabular}
\caption{Continuum properties in the phase interval {\em\bf 0,4}-{\em\bf 0,8}}
\label{tab2}
\end{table}

\begin{figure}[t]
\resizebox{\hsize}{!}{\includegraphics{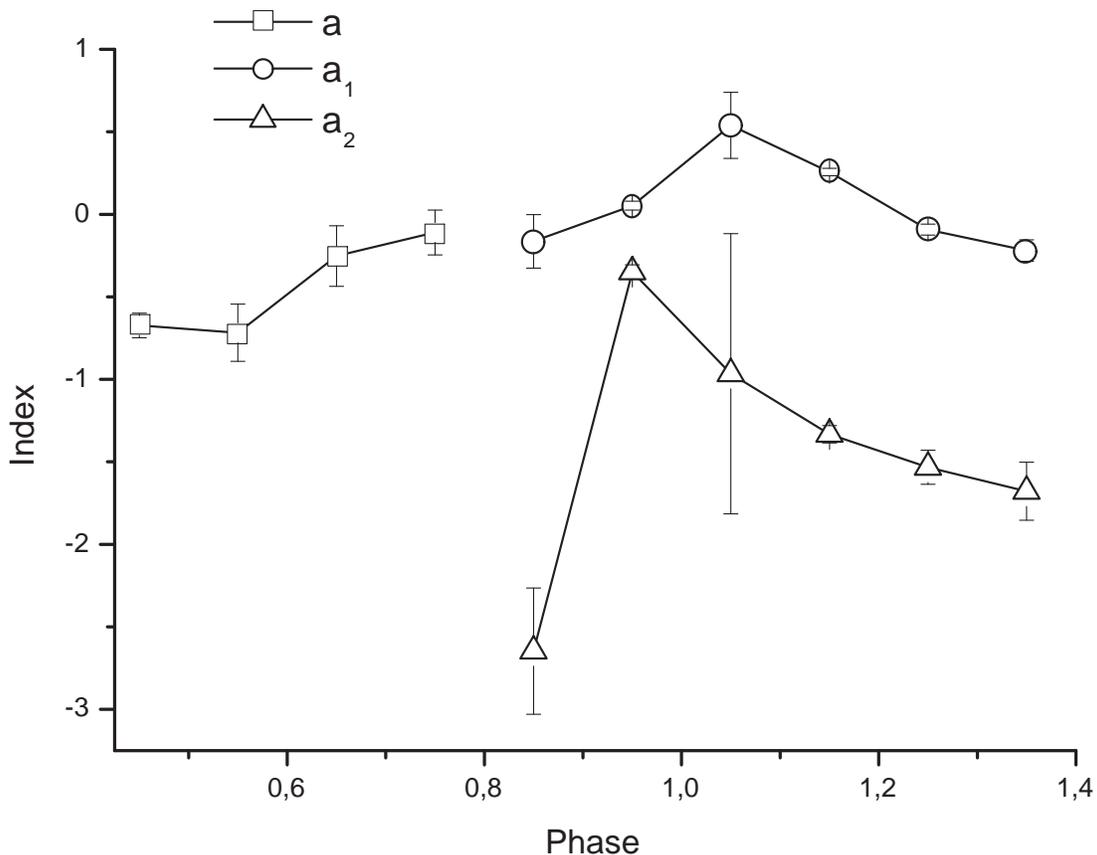}}
\caption{Power-law indexes. Detail description of the parameters $a$,$a_1$ and $a_2$ see on the formulas (\ref{shape}) and (\ref{shape2}).}
\label{fig4}
\end{figure}

\begin{figure}[t]
\resizebox{\hsize}{!}{\includegraphics{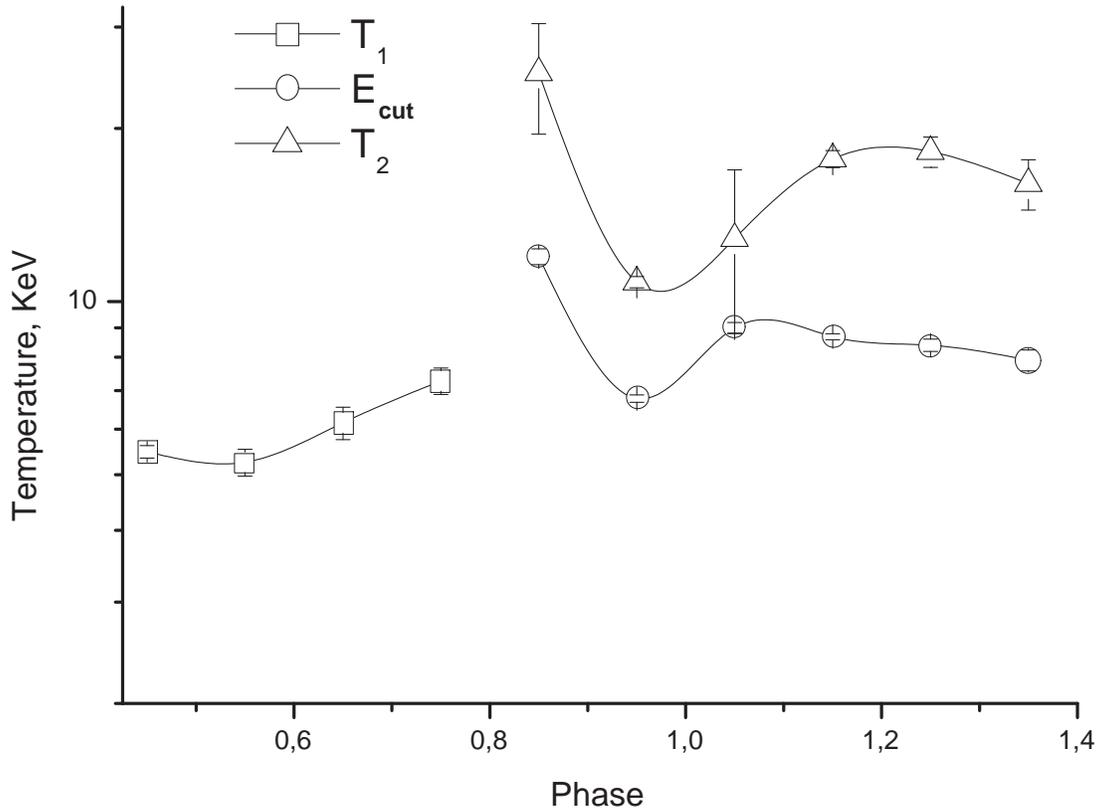}}
\caption{Temperature $T_1$ and $T_2$, and cutoff energy $E_{cut}$.
 Detail description of the parameters $T_1$,$T_2$ and $E_{cut}$ see on the formulas (\ref{shape}) and (\ref{shape2}).}
\label{fig5}
\end{figure}

\begin{table}[t]

\begin{tabular}{cr@{$\pm$}lr@{$\pm$}lr@{$\pm$}lr@{$\pm$}l}
\hline
\hline
Phase & \multicolumn{2}{c}{$a_1$}& \multicolumn{2}{c}{$a_2$}& \multicolumn{2}{c}{$E_{cut}$}& \multicolumn{2}{c}{$T_2$}\\
\hline
0,85 & -0,164 & 0,16 & -2,65 & 0,38 & 12,0 & 0,38 & 25,0 & 5,5\\
0,95 & 0,0525 & 0,027 & -0,349 & 0,043 & 6,78 & 0,10 & 10,8 & 0,24\\
1,05 & 0,539 & 0,20 & -0,966 & 0,85 & 9,00 & 0,19 & 12,9 & 4,1\\
1,15 & 0,256 & 0,023 & -1,33 & 0,052 & 8,69 & 0,10 & 17,7 & 0,61\\
1,25 & -0,0926 & 0,033 & -1,53 & 0,10 & 8,40 & 0,21 & 18,2 & 1,1\\
1,35 & -0,219 & 0,065 & -1,68 & 0,18 & 7,91 & 0,33 & 16,0 & 1,6\\
\hline
\end{tabular}
\caption{Continuum properties in the phase interval {\em\bf 0,8}-{\em\bf 1,4}}
\label{tab3}
\end{table}

\begin{figure}[t]
\resizebox{\hsize}{!}{\includegraphics{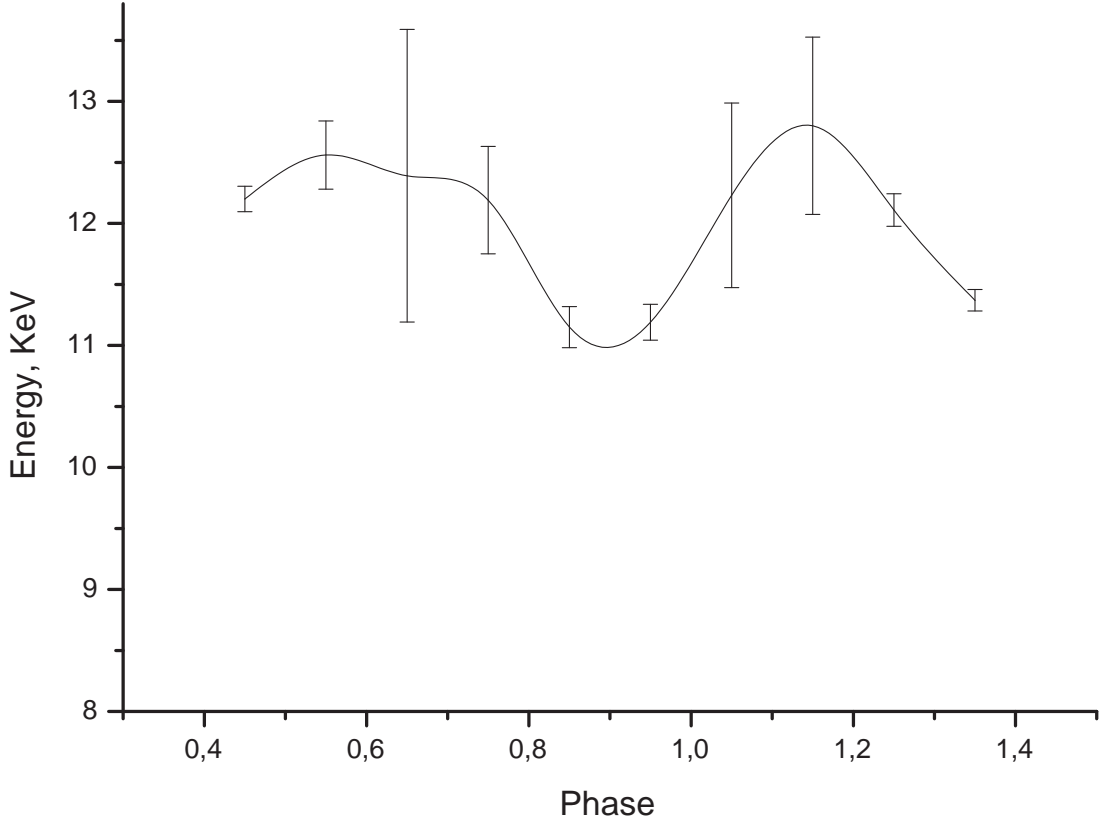}}
\caption{Energy of the first cyclotron harmonic $E_1$.}
\label{fig6}
\end{figure}

\begin{figure}[t]
\resizebox{\hsize}{!}{\includegraphics{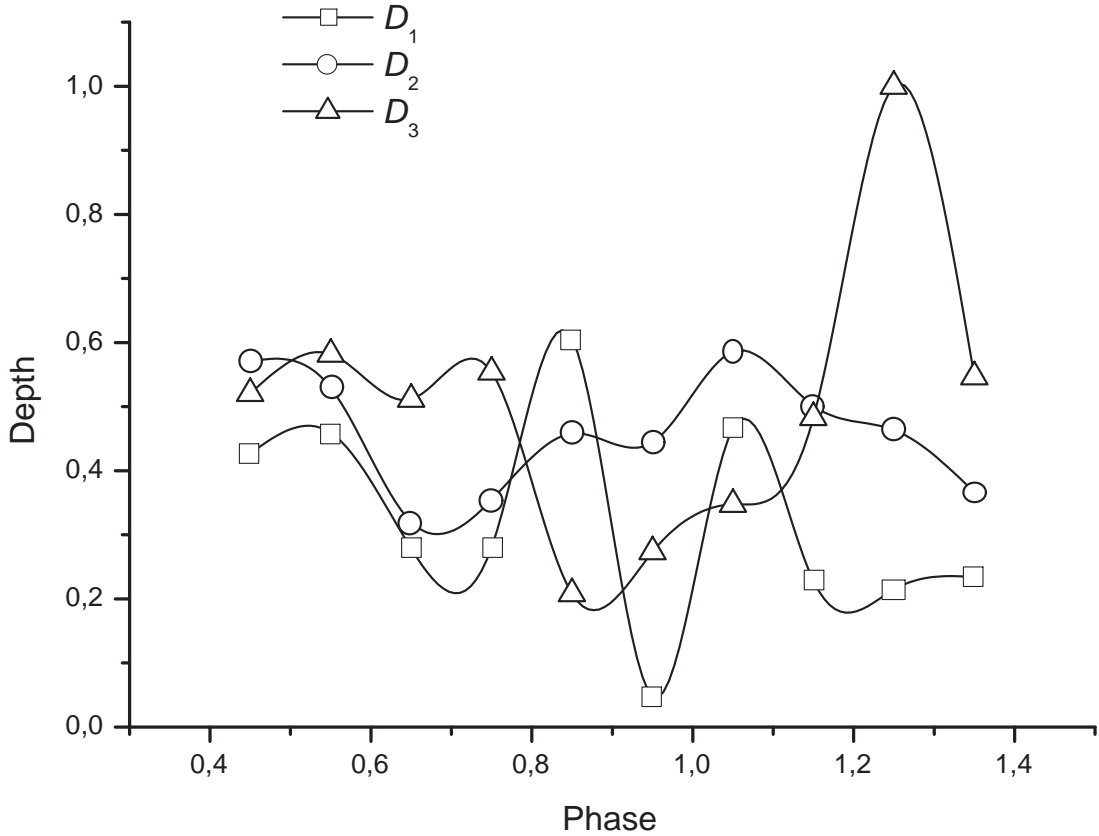}}
\caption{Depths of the cyclotron lines.}
\label{fig71}
\end{figure}

\begin{table}[p]
\resizebox{0.7\hsize}{!}{
\begin{tabular}{cr@{$\pm$}lr@{$\pm$}lr@{$\pm$}lc}
\hline
\hline
Phase & \multicolumn{2}{c}{$D_1$}& \multicolumn{2}{c}{$D_2$}& \multicolumn{2}{c}{$D_3$}& $D_4$ \\
\hline
0,45 & 0,43 & 0,031 & 0,57 & 0,049 & 0,52 & 0,041 & ---  \\
0,55 & 0,46 & 0,091 & 0,53 & 0,055 & 0,58 & 0,063& ---  \\
0,65 & 0,28 & 0,25 & 0,32 & 0,10 & 0,51 & 0,070& --- \\
0,75 & 0,28 & 0,095 & 0,35 & 0,042 & 0,55 & 0,045& --- \\
0,85 & 0,61 & 0,083 & 0,46 & 0,061 & 0,21 & 0,050& --- \\
0,95 & 0,048 & 0,0090 & 0,44 & 0,017 & 0,27 & 0,041 & 0,32$\pm$0,025\\
1,05 & 0,47 & 0,14 & 0,59 & 0,087 & 0,35 & 0,043 & 0,43$\pm$0,25\\
1,15 & 0,23 & 0,0095 & 0,50 & 0,0071 & 0,48 & 0,37 & 0,26$\pm$0,033\\
1,25 & 0,21 & 0,020 & 0,46 & 0,010 & 1.0 & 3,6& --- \\
1,35 & 0,23 & 0,026 & 0,37 & 0,016 & 0,55 & 0,37& --- \\
\hline
\end{tabular}}
\caption{Depths of the cyclotron lines.}
\label{tab11}
\end{table}

\begin{figure}[t]
\resizebox{\hsize}{!}{\includegraphics{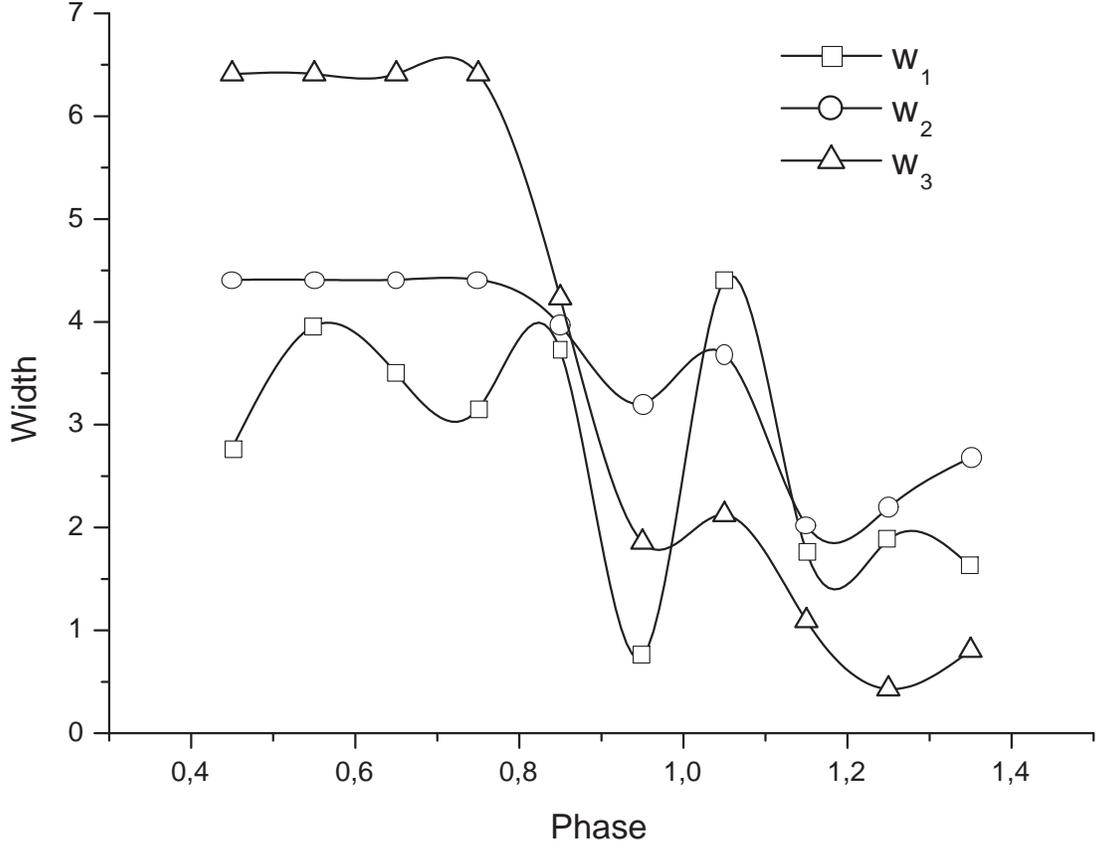}}
\caption{Widths of the cyclotron lines.}
\label{fig72}
\end{figure}

\begin{table}[p]
\resizebox{0.7\hsize}{!}{
\begin{tabular}{cr@{$\pm$}lr@{$\pm$}lr@{$\pm$}lc}
\hline
\hline
Phase & \multicolumn{2}{c}{$w_1$}& \multicolumn{2}{c}{$w_2$}& \multicolumn{2}{c}{$w_3$}& $w_4$ \\
\hline
0,45 & 2,77 & 0,21 & 4,41 & 2,1 & 6,41 & 5,8 & --- \\
0,55 & 3,96 & 1,1 & 4,41 & 3,0 & 6,41 & 6,9 & --- \\
0,65 & 3,50 & 1,4 & 4,41 & 8,6 & 6,41 & 7,4 & --- \\
0,75 & 3,16 & 0,58 & 4,41 & 3,4 & 6,41 & 4,8 & --- \\
0,85 & 3,72 & 0,66 & 3,97 & 0,51 & 4,23 & 1,4 & --- \\
0,95 & 0,749 & 0,22 & 3,21 & 0,14 & 1,86 & 0,60 & 8,98$\pm$1,9\\
1,05 & 4,41 & 4,04 & 3,69 & 0,68 & 2,12 & 0,58 & 12,7$\pm$5,1\\
1,15 & 1,77 & 0,13 & 2,02 & 0,065 & 1,09 & 0,97 & 7,69$\pm$1,9\\
1,25 & 1,88 & 0,20 & 2,19 & 0,099 & 0,429 & 0,77 & --- \\
1,35 & 1,62 & 0,11 & 2,68 & 0,19 & 0,805 & 0,64 & --- \\
\hline
\end{tabular}}
\caption{Width of the cyclotron lines.}
\label{tab12}
\end{table}

\begin{figure}[t]
\resizebox{\hsize}{!}{\includegraphics{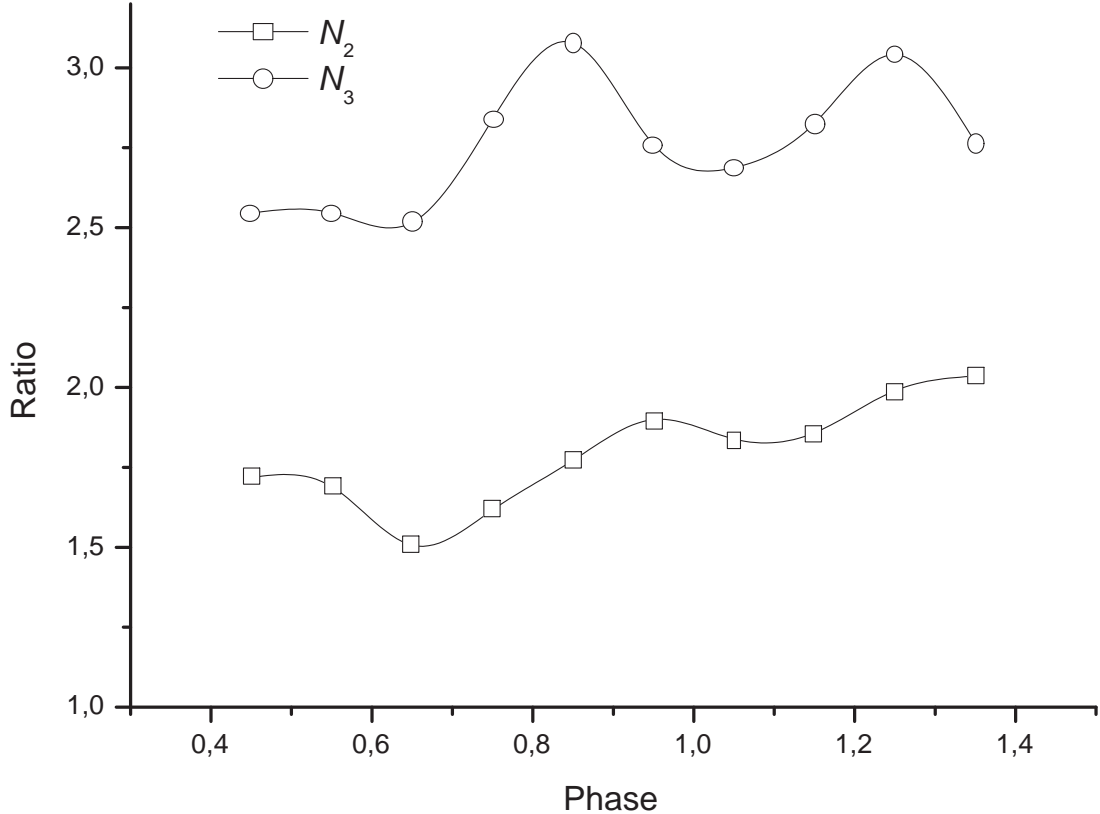}}
\caption{Positions of the cyclotron lines. See the significance
of the parameters $N_2$ and $N_3$ in the formula (\ref{gauss}).}
\label{fig73}
\end{figure}

\begin{table}[p]
\resizebox{0.7\hsize}{!}{
\begin{tabular}{cr@{$\pm$}lr@{$\pm$}lr@{$\pm$}lc}
\hline
\hline
Phase & \multicolumn{2}{c}{$E_1$}& \multicolumn{2}{c}{$N_2$}& \multicolumn{2}{c}{$N_3$}& $N_4$ \\
\hline
0,45 & 12,2 & 0,10 & 1,72 & 0,038 & 2,55 & 0,092 & ---\\
0,55 & 12,56 & 0,28 & 1,69 & 0,052 & 2,55 & 0,11 & ---\\
0,65 & 12,39 & 1,2 & 1,51 & 0,056 & 2,52 & 0,30 & ---\\
0,75 & 12,19 & 0,44 & 1,62 & 0,034 & 2,84 & 0,12 & ---\\
0,85 & 11,15 & 0,17 & 1,77 & 0,020 & 3,08 & 0,089 & ---\\
0,95 & 11,19 & 0,15 & 1,9 & 0,027 & 2,76 & 0,046 & 3,77$\pm$0,11\\
1,05 & 12,23 & 0,76 & 1,84 & 0,12 & 2,69 & 0,17 & 3,79$\pm$0,34\\
1,15 & 12,8 & 0,73 & 1,86 & 0,0099 & 2,83 & 0,035 & 4,16$\pm$0,10\\
1,25 & 12,11 & 0,13 & 1,99 & 0,020 & 3,04 & 0,053 & ---\\
1,35 & 11,37 & 0,088 & 2,04 & 0,015 & 2,77 & 0,034 & ---\\
\hline
\end{tabular}}
\caption{Fundamental cyclotron energy and positions of the cyclotron lines.}
\label{tab13}
\end{table}

In order to obtain the energy-resolved pulse-profiles and
phase-resolved energy spectra the photon arrival times were
first corrected to the Solar System barycenter. By using period
folding \cite{leahy83}, we have found the period of the pulsar to
be equal $3.6143\pm 0.0002$~s.
 No variations of the period of the source have been detected.
Obtained pulse profiles for different energy ranges are
represented on Fig.\ref{fig1a}-\ref{fig1c}. A  main soft peak
(in the phase interval
 {\em\bf 0,2}-{\em\bf 0,75}) with maximum at {\em\bf 0,4} and a hard peak (in the phase interval
 {\em\bf 0,75}-{\em\bf 1,2}) with maximum at {\em\bf 1,05} are
plainly distinguishable. Hard peak is relatively narrow, and its shape remains almost
constant over a wide range of energies (above 4~keV). On the
contrary, the form of the main soft peak depends highly from energy;
over 30~{keV} it almost disappears.

To examine the situation in more details we plotted the softest
part of the emission with higher energy resolution
(Fig.\ref{softdetail}) .On this figure a new feature appears:
a secondary soft peak with maximum at the phase {\em\bf 0,85},
while the hard peak disappears. Moreover, the shape of the main
soft peak changes drastically. On the energies above 3~keV it has
a clear maximum at the phase {\em\bf 0,4}. and descending region
between {\em\bf 0,4} and {\em\bf 0,7}. But over the energies
2-3~keV it becomes quite flat without any clear maximum. It might
be accounted for the influence of the above-mentioned secondary
soft peak, but if we consider phase profile evolution from 5 to
2~keV, it turns out, that intensity difference between the maximum
value at the phase {\em\bf 0,85} and the minimum value at
$\sim${\em\bf{0,75}} remains almost the same, while intensity
difference between the minimum value at {\em\bf 0,75} and the
value at {\em\bf 0,7} rises steeply. It means that the appearance
of the secondary soft peak and the change of the shape of the main
soft peak at low energies are two different phenomena.

\subsection{Spectral analysis.}
In order to perform a phase-resolved analysis of the spectrum of
the pulsar, all phase interval has been divided on ten equal phase
regions. The phase region selection is represented on
Fig.\ref{fig09}.

The source 4U0115+63 has a power-law spectrum with high-energy
cutoff; this continuum is strongly distorted by cyclotron
absorption \cite{white83, santangelo99b}. It turned out necessary
during our analysis to divide the whole phase interval on two
regions and to use different continuum models for them.  First
region covers phase interval from {\em\bf 0,4} up to {\em\bf 0,8}
and approximately corresponds to the descending edge of a lower
peak of the pulse-profile; second region covers all  other
phases. For the first region
 it was enough to use a simple power-law with exponential cutoff
model

\begin{equation}
\label{shape}
 f(E)\propto E^{\displaystyle{a}}
\exp\left(\displaystyle{-\frac{E}{T_1}}\right)
\end{equation}

 For the second region
 it turned out to be insufficient. The continuum
spectra within the main peak have been fitted by the following
model:

\begin{equation}
\label{shape2}
f(E)=\left\{
\begin{array}{rl}
A E^{\displaystyle{a_1}},&E<E_{cut}\\
B E^{\displaystyle{a_2}} \exp\left(\displaystyle{-\frac{E}{T_2}}\right),&E>E_{cut}\\
\end{array}
\right.
\end{equation}
Numerical coefficients $A$ and $B$ have been chosen so that to
make the function continuous. A fracture on the energy $E_{cut}$
has been smoothed by a polynomial function.

In order to represent the cyclotron feature these continuum models
have been multiplied on the so-called Gaussian filters, one for
each line.
\begin{equation}
\label{gauss}
G_n(E)=1-D_n \exp\left(-\frac{(E- N_n \cdot E_1)^2}{2w_n^2}\right)
\end{equation}
Here $n$ is the number of harmonic, $E_1$ is an energy of the
first harmonic. Parameters $N$ define positions of the other
harmonics. Initially they were taken equal to $n$, but they were
considered as free parameters in order to represent unequidistant
cyclotron lines. Three cyclotron lines have been detected at all
phases. The fourth harmonic turned out to be necessary only at the
phases {\em\bf 0,9-1,2}. No iron line at 5-7~{keV} has been
detected.

Described model turned out to be as a whole quite acceptable to
fit the spectra, though in some phase intervals certain problems
appeared (see Discussion). Results, obtained with the model, are
represented on tables \ref{tab2}-\ref{tab13} and figures
\ref{fig4}-\ref{fig73}.

\section{Results and discussion.}
\subsection{Energy-resolved  analysis of the phase profiles.}
As it has been described in the previous section, pulse profile
of the source contains two main features: the hard peak and the
main soft peak. This brings up a question: Are these features
produced by different magnetic poles of the neutron star, or they
are a pencil and a fan beam from the same pole. Let us discuss
both these hypotheses. But at first we remark, that if the
emission has a fan directional diagram, an observer should cross
its plane twice a period. Therefore we can expect a fan beam
should produce two features on a pulse profile. And so the hard
peak is the most probably a pencil beam (there is only one hard
peak in the pulse profile), while the two peaks could either be
produced by the same fan beam (the phase distance between them is
almost equal {\em\bf 0,5}), or be two independent components,
produced by different magnetic poles.

If the hard peak and the main soft peak are formed in the same
active region (this supposition is called hereafter {\em\it hypothesis 1}), then the
main soft peak is, of cause, a fan beam (the phase interval between it and the hard peak is big enough), and
the secondary soft peak is almost indubitably the second impulse
of this fan. The hard and the main soft peaks together cover
nearly the whole period of the pulsar. It means that one active zone of the neutron
star is visible all the period, while an emission of the
other does not reach an observer at all. This situation is
possible only if the angle between our line of sight and the
axis of rotation of the neutron star is small enough.

The hard peak is situated strongly asymmetrically with respect to
the two soft peaks, which is to say that the pencil beam is
inclined noticeably with respect to the axis of the fan beam.
Moreover, the spectral properties of the main and the secondary
soft peaks are quite different, i.e spectral characteristics of
the fan beam depend highly from azimuth. Surely, it is possible
only in the accretion occurs non-axis-symmetrically. It means,
that the magnetic field of the pulsar is greatly non-dipole.

Let us now consider the situation, when the hard peak and the main
soft peak are formed in different active regions (hereafter
{\em\it hypothesis 2}). Then the main soft peak and the secondary
soft peak could be components of a fan beam forming on the second
magnetic pole of the neutron star ({\em\it hypothesis 2a}). In
conclusion, the situation is possible when the main soft peak is
formed by a pencil beam from the second magnetic pole, while the
secondary soft peak is only a softer and wider component of the
hard peak ({\em\it hypothesis 2b}). But in any case there are
strong arguments that the field of the pulsar differs
drastically from the dipole one. The hard peak is situated
strongly asymmetrically with respect to the two soft peaks (this
argument is valid if we assume {\em\it hypothesis 2a}) and the
phase interval between the hard peak and the main soft peak is
well below $0.5$ (if we assume {\em\it hypothesis 2b}).

Under consideration of the {\em\it hypothesis 2} a question
appears: why we can not see the hard component from the second
magnetic pole? There are two probable explanations of this fact.
As the hard component is more concentrated, it is possible that
the geometry of the source is so that a hard component from the
second hot spot on the neutron star's surface (corresponding to
the soft peak) exists but is invisible for us. But it is also
possible, that emission of this hot spot doesn't contain a hard
component at all. As it has been already mentioned, magnetic
field of the pulsar differs strongly from the dipole one. Under
such conditions the physical parameters of the accreting streams
and, consequently, of the emitting zones are absolutely different
for the magnetic poles of the neutron star. It can led to the
absence of a hard component from one of the poles.

Both hypotheses ({\em\it hypothesis 1} and {\em\it hypothesis 2})
have arguments for and against. If we assume that the hard peak
and the main soft peak are formed on different active regions,
then (considering the above-mentioned departure of the magnetic
field from the dipole one) we might expect that the cyclotron
features in the spectra of the hard and the main soft peaks have
different fundamental energies. As seen from Fig.\ref{fig6},
this is not the case. Changes of the first cyclotron line's
energy doesn't correlate with the peaks of emission. This
testifies in favour of formation of all the emission in the same
region ({\em\it hypothesis 1}). But the {\em\it hypothesis 1}
faces with an another trouble. As it will be discussed below, the
emission of the pulsar likely appears as a result of
comptonization of the emission by hot plasma. Spectral properties
of the obtained emission depends upon the temperature of the
plasma \cite{suntit80}. The secondary peak disappears already at
3-4~keV, the main soft peak -- at 25-30~keV; The hard peak is
visible up to the highest energies. If we assume the {\em\it
hypothesis 1}, it means that in a single active zone there are
tree regions of plasma with absolutely different temperatures,
which on the one hand have an optical depths great enough to
produce remarkable features on the pulse profile (and each
feature has their own directional characteristic), but on the
other hand each region does not influence at all on the emission
of the others. This situation is, of cause, possible, but it seems
a bit artificial. It would be more natural to suppose, that the
emission is formed in two different zones with different physical
conditions.

\subsection{Spectral analysis.}
The source 4U0115+63 has a power-law spectrum with high-energy
cutoff . This shape of spectrum as a whole is typical for the
X-ray binaries and likely appears as a result of comptonization of
the emission by the hot plasma of the accretion column. We tried
to fit the continuum by the spectra suggested in
\cite{suntit80}. These spectra represent results of
comptonization by a simple-geometry plasma slab. Our attempts
failed. It means that the geometry and the temperature
distribution in the active region of the source are rather
complicate and they probably consist of several layers of plasma
with various temperatures and optical depths.

The cyclotron lines exist definitely at all the phases of the
pulsar, but parameters of the spectrum are very phase-dependable.
The parameters are represented on tables \ref{tab2}-\ref{tab3}.
First, second and third harmonics have been detected at all the
phases of the source; fourth harmonic was visible only in the hard
peak.

The cyclotron feature is the most pronounced not on the maxima of
luminosity but on the descending edges. Moreover, the fundamental
cyclotron energy also has a maximum on these edges. One of the
possible ways of explanation of this effect is the following: an
emission propagating exactly along the accretion column interacts
with plasma on a very large interval of heights. The main energy
of a cyclotron line is formed well above the surface of the
neutron star in a region with respectively weak magnetic field.
On the contrary an emission propagating aslant with reference to
the column leaves the column respectively deep, but its path in
these deeper layers is longer. Consequently, we can expect that in
this case that cyclotron line would be deeper but narrower and its
main energy higher. One can see that it is really so: the widths
of the lines have local maxima correlating with the maximum of
luminosity instead of the descending edge. But then a question
appears: why this effect acts only on the descending edge and not
valid on the ascending one? The cause of this could be the
above-mentioned asymmetry of the accretion column appearing as a
result of out-of-dipole magnetic field of the pulsar. The
difference between the spectra of descending and ascending edges
can be considered in its turn as an addition evidence of a
departure of the field from a dipole.

The fundamental cyclotron energy is not stable: its variations are
of the order of 10\% and they are near correlating with the
variations of luminosity. Harmonics are very unequidistant. This
effect is the most distinct on the second harmonic. The ratio
between energies of the second and the first harmonics is as
small as 1.6 at some phases.

The situation with the theoretical interpretation of these effects
is not so simple. The effects are usually interpreted as a result
of complex geometry of the emission forming region and magnetic
field variations in it. Let as analyze this possibility.

The neutron star is considered to have a radius $\sim$10~{km} and
a dipole magnetic field (for the estimating consideration this
supposition is quite acceptable). The luminosity of the source
4U0115+63 is $\sim 10^{37}~\mbox{erg/sec}$. When determined from
the cyclotron energy by the usual nonrelativistic formula
\begin{equation}
\label{cyc}
E_1=\hbar\frac{e H}{m_e c}
\end{equation}
the magnetic field of the source appears to be $1.1\cdot
10^{12}~\mbox{Gs}$. There is a well-known formula for the angular
size $\chi$ of the hot spot on the X-ray pulsar's surface if its
magnetic field is a dipole.
\begin{equation}
\label{spot}
\chi \simeq \sin \chi = 7.5 \cdot 10^{-4} H^{\textstyle -\frac{2}{7}}
L^{\textstyle \frac{1}{7}}
\end{equation}
In the considering case we obtain $\chi \simeq 0.06$. Relative
variation of the magnetic field (and hence of the cyclotron
energy) within the boundaries of such small a spot is of the order
of ${\frac{\Delta H}{H}} \approx 1.4 \cdot 10^{-3}$. Of course it
is enough to explain neither widths of lines nor their arranging.
Consequently all these effects should be considered as a result of
variations of the magnetic field along the height of the
accretion column.

In the average the second harmonic is the strongest. The first
harmonic is the most pronounced only in the region out of the
major peak (phases {\em\bf 0,45}-{\em\bf 0,85}). At the phase
{\em\bf 1,15} where the cyclotron feature is the most distinct the
second and the third harmonics have almost the same depths which
is significantly higher then the depths of the first and the
fourth one. These ratios between depths of different harmonics
looks strangely. As it was shown in \cite{harding91}, cyclotron
cross sections averaged over directions and polarizations under
considering physical conditions are related as
\begin{equation}
\label{kand1}
\sigma_1 :\sigma_2 :\sigma_3 = 400:20:3
\end{equation}
Even for the direction perpendicular to the magnetic field this
ratio is
\begin{equation}
\label{kand2}
\sigma_1 :\sigma_2 :\sigma_3 = 200:18:3
\end{equation}
One can see that in any case the first harmonic should be
significantly stronger then the others. Meanwhile at the main
peak the second and the third harmonics are stronger then the
first one. The most believable explanation of this fact was
given in the series of articles (see \cite{wang89physrev},
\cite{wang89}, \cite {lamb90}, \cite{wang93},
\cite{isenberg98}). The authors explain the contradiction by
the influence of the Raman scattering. Under the physical
conditions existing in the accretion column the probability of the
transition of an electron from the second Landau level through the
branch ($2\to 1\to 0$) is approximately seven times higher
\cite{daugherty77} then through the branch ($2\to 0$). An
influence of the true absorption is negligible
\cite{bonazzola79}. As a result a photon of the second harmonic
being absorbed decays on two photons of the first one. This
spawning leads to the significant decreasing of the first
harmonic's depth.

But this mechanism could be acceptable only for the second
harmonic: for the ratio between third and second harmonics it
fails. In fact, the integrated probability of the branches ($3\to
1\to 0$) and ($3\to 2\to 0$) is only 29\%, and only one photon of
the second harmonic appears instead of two photons in the
previously described case. Furthermore even this rare photon has
no chance to reach the distant observer. As it was shown in
\cite{lamb90}, Raman scattering of photons of the third harmonic
can influence effectively on the spectra of previous harmonics
only if an optical depth of the scattering electrons on the third
harmonic is at least comparable with 1. But in this case their
optical depth on the second harmonic should be $\sim$6. The
spawned photon of the second harmonic is sure to experience the
second scattering and decay on two photons of the first
harmonic. So the efficiency of the spawning of photons of
harmonics of the orders higher then first is negligible. But
ratios of their depths in the observing spectrum are also very far
from the theoretical relations between cyclotron cross-sections.

The nonequidistancy of the harmonics is also very difficult for
the theoretical explanation. Owing to the above-mentioned
relations between the cyclotron cross-sections the magnetized
substance is much more transparent on the energy of the second
harmonic then on the energy of the first one. Consequently the
second harmonic should appear in the deeper regions of the
accretion column with the stronger magnetic field. We might
expect the distance between first and second harmonics to be
bigger then the harmonic one. Meantime the distance is smaller
than the harmonic: the ratio between energies of the second and
the first harmonics sometimes is as small as $1.6$. The same
result have been already reported in \cite{heindl99}.

The third harmonic is grotesquely deep at the phase {\em\bf 1,35}.
Moreover the confidence intervals of the $D_3$ at phases {\em\bf
1,25-1,45} are too big. The reasons of this effect is that the
real line shapes differ distinctly from the Gaussian. The same
result has been already reported \cite{mihara04}. Unfortunately
there is now no universally accepted view of what shape should
have a cyclotron line in a spectrum of an X-ray pulsar. A Gaussian
model is one of the most commonly-using, mainly because of its
simplicity. We tried different kinds of models for the lines and
the Gaussian turned out to be the best. Moreover, the emission at
the phase {\em\bf 1,25} is a mixture of emissions of the hard peak
and the main soft peak and it makes its spectrum very complex. One of the possible ways
to avoid this difficulties is to suppose that the cyclotron feature is formed by
cyclotron radiation of ultrarelativistic, anisotropic electrons \cite{baushev02}.

\section{Conclusion.}
To summarize: We studied the 4-75~{keV} pulse phase resolved
spectrum of 4U0115+63 observed by Beppo-Sax. Analysis of it
suggests that the magnetic field of the neutron star differs
drastically from the dipole one. We confirm the strong
dependence of the parameters of the cyclotron feature in the
spectrum on the phase, while the continuum part changes rather
weakly. We also confirm the existence of the fourth cyclotron
line near $\sim 45$~{keV}. Gaussian shape is not quite good to
fit the lines; further investigations are needed to propose some
better model.

Consideration of the physical processes of the cyclotron feature
formation points to some serious difficulties of the absorption
model. In particular, it is unclear how to explain the observing
ratios between depths and central energies of cyclotron lines in
the context of it. Thus the generally used absorption model of the
cyclotron feature in the spectrum of the source can not be
considered as an indisputable.

\section{Acknowledgments}
We acknowledge with deep gratitude the assistance we have received from our colleagues
from IASF, especially from A. La Barbera, A. Santangelo, and A. Segreto.


\begin{thebibliography}{99}

\bibitem{rappaport78}
Rappaport, S., Clark, G.W., Cominsky, L. et al.,
Astrophys.J. Lett. {\bf 224}, 1 (1978).

\bibitem{unger98}
Unger, S., Roche, P., Negueruela, I., et al. 1998,  A\&A {\bf 336}, 960 (1998).

\bibitem{forman76}
Forman, W., Jones, C. and Tananbaum, H.,
Astrophys.J. Lett. {\bf 206}, 29 (1976).

\bibitem{rose79}
Rose, L., Pravdo, S., Kaluzienski, L. et al., Astrophys.J. {\bf 231}, 919 (1979).

\bibitem{bildsten97}
Bildsten, L. et al., Astrophys.J.Supp. {\bf 113}, 367 (1997).

\bibitem{coburn02}
Coburn, W., Heindl, W.A., Rothschild, R.E. et al. 2002,
Astrophys.J. {\bf 580}, 394 (2002).

\bibitem{mihara04}
Mihara, T., Makishima, K., Nagase, F.,
Astrophys.J. {\bf 610}, 390 (2004).

\bibitem{santangelo99b}
Santangelo, A., Segreto, A., Giarrusso, S. et al.,
Astrophys.J. Lett. {\bf 523}, 85 (1999).

\bibitem{wheaton79}
Wheaton, W.A. et al., Nature {\bf 282}, 240 (1979).

\bibitem{white83}
White, N. E., Swank, J. H. and Holt, S. S., Astrophys.J.
{\bf 270}, 711 (1983).

\bibitem{heindl99}
Heindl, W.A., Coburn, W., Gruber, D. E. et al., Astrophys.J. Lett.
{\bf 521}, 49 (1999).

\bibitem{boella97}
Boella, G., et. al., A\&A {\bf 122}, 327 (1997).

\bibitem{manzo97}
Manzo, G., Giarrusso, S., Santangelo, A., et al., A\&A {\bf 122}, 341 (1997).

\bibitem{frontera97}
Frontera, F., et al., A\&A {\bf 122}, 357 (1997).

\bibitem{leahy83}
Leahy, D.A., Elsner, R.F., Weisskopf, M.C.
Astrophys.J. Lett. {\bf 272}, 256 (1983).

\bibitem{suntit80}
Sunyaev, R.A., Titarchuk, L.G. 1980, A\&A
{\bf 86}, 121.

\bibitem{baushev02}
Baushev, A.N. 2002, Astronomy Reports, {\bf 46}, 830; arXiv:0804.1592

\bibitem{harding91}
Harding, A.K., Daugherty, J.K., Astrophys.J. {\bf 374},
687 (1991).

\bibitem{wang89physrev}
Wang, J.C.L., et al.  Phys. Rev. Lett. {\bf 63},
1550 (1989).

\bibitem{wang89}
Wang, J.C.L., Wasserman, I.M. and Salpeter, E.E.,
Astrophys.J. {\bf 338}, 343 (1989).

\bibitem{lamb90}
Lamb, D.Q., Wang, J.C.L. and Wasserman, I.M., Astrophys.J. {\bf 363}, 670 (1990).

\bibitem{wang93}
Wang, J.C.L., Wasserman, I.M. and Lamb, D.Q.,  Astrophys.J. {\bf 414}, 815 (1993).

\bibitem{isenberg98}
Isenberg, M., Lamb, D.Q. and Wang, J.C.L., Astrophys.J. {\bf
505}, 688 (1998).

\bibitem{daugherty77}
Daugherty, J.K., Ventura, J., A\&A {\bf 61}, 723 (1977).

\bibitem{bonazzola79}
Bonazzola, S., Heyvaerts, J. and Puget, J.L., A\&A {\bf
78}, 53 (1979).

\end{thebibliography}
\end{document}